# Constraining The Hubble Parameter Using Distance Modulus – Redshift Relation

*C. C. Onuchukwu*[1], A. C. Ezeribe[2]

[1.] Department of Industrial Physics, Anambra State University, Uli (email:onuchukwu71chika@yahoo.com)
[2.] Department of Physics/Industrial Physics, Nnamdi Azikiwe University, Awka (ac.ezeribe@unizik.edu.ng)

**ABSTRACT:**

Using the relation between distance modulus ($m - M$) and redshift ($z$), deduced from Friedman-Robertson-Walker (FRW) metric and assuming different values of deceleration parameter ($q_0$). We constrained the Hubble parameter ($h$). The estimates of the Hubble parameters we obtained using the median values of the data obtained from NASA Extragalactic Database (NED), are: $h = 0.7 \pm 0.3$ for $q_0 = 0$, $h = 0.6 \pm 0.3$, for $q_0 = 1$ and $h = 0.8 \pm 0.3$, for $q_0 = -1$. The corresponding age ($\tau$) and size ($\Re$) of the observable universe were also estimated as: $\tau = 15 \pm 1\ Gyrs$, $\Re = (5 \pm 2) \times 10^3\ Mpc$, $\tau = 18 \pm 1\ Gyrs$, $\Re = (6 \pm 2) \times 10^3\ Mpc$ and $\tau = 13 \pm 1\ Gyrs$, $\Re = (4 \pm 2) \times 10^3\ Mpc$ for $q_0 = 0$, $q_0 = 1$ and $q_0 = -1$ respectively.

## 1  INTRODUCTION

The universe is all of space, time, matter and energy that exist. The ultimate fate of the universe is determined through its gravity, thus, the amount of matter/energy in the universe is therefore of considerable importance in cosmology. The Wilkinson Microwave Anisotropy Probe (WMAP) in collaboration with National Aeronautics and Space Administration (NASA) from one of their mission estimated that the universe comprises of about 4.6% of visible matter, about 24% are matter with gravity but do not emit observable light (the dark matter) and about 71.4% was attributed to dark energy (possibly anti-gravity) that may be responsible for driving the acceleration of the observed expansion of the universe (Reid et al. 2013). The universe has been observed to be expanding, first suggested by Einstein in his general theory of relativity and subsequently observed (e.g. Hubble 1929; Spergel et al. 1997, 2003; Riess et al. 2011).

Amongst the different models proposed to explain the evolution of the universe, the Big Bang Theory is presently the most acceptable model that described most of the observational features in the evolution of the universe. It explained that the universe may have started from an initial singularity (an extremely hot dense phase generally refer to as the Planck epoch). At this stage all the four fundamental forces were unified and absolute symmetry is assumed. At the end of the Planck epoch, gravitational force separated from the gauge forces. Other era usually considered in Big Bang Theory (with the associated symmetry breaking) include in the Grand Unification Epoch and Inflationary Epoch (at the end of which the strong forces separated from the electroweak force), the Electroweak Epoch (when there was unification of electromagnetism and weak nuclear interaction), epoch of Baryogenesis, Hadron Epoch, Lepton Epoch, Nucleosynthesis, Photon epoch, Recombination Epoch and this present time, the Matter Dominated Era (Robson 1996; Bonometto et al. 2002; Roos 2003; Harrison 2003; Ryden 2003).

The generally acceptable mathematical theory for studying the evolution of the universe is general relativity (Einstein 1916). In a uniform universe, general relativity has a simple solution for the evolution of the geometry of the universe (contraction or expansion) which depends on its content and past history (Spergel et al. 1997). In the presence of enough matter, the expansion is expected to slow down or even change to contraction. On the other hand, the dark energy (which can be represented the cosmological constant ($\Lambda$)) may be driving the universe towards increasing expansion (Spergel et al. 1997). The current rate of expansion is usually expressed by the Hubble constant ($H_0$) with an estimated value of $100h\ kms^{-1}\ Mpc^{-1}$, with the Hubble parameter ($h$) having values of $h = 0.5 - 1.0$ (Roos 2003; Ryden 2006).

One of the fundamental goals of cosmology is to determine the expansion rate of the universe. Various methods that have been applied include:

i. The use of distant Type 1a supernovae (SN 1a) as standard candles (Riess et al. 1998; Amanuallah et al. 2010; Conley & Sullivan 2011; Suzuki et al. 2011). The apparent peak magnitude of these supernovae yields a relative luminosity distance $d_L$ as a function of redshift from which the Hubble constant is estimated (Wang & Tegmark 2005; Sollerman et al. 2009; Shafieloo & Clarkson 2010).

ii. Large galaxy surveys for mapping of cosmic distances and expansion – by using the large scale clustering pattern of galaxies which contain the signature of Baryon Acoustic Oscillation (BAOs). BAO refers to the regular periodic fluctuations in the density of the visible baryonic matter of the universe, caused by acoustic waves which existed in the early universe. By looking at large scale clustering of galaxies, a preferred length-scale which was imprinted in the distribution of photons and baryons propagated by the sound waves in the relativistic plasma of the early universe, can be calibrated by the observation of Cosmic Microwave Background Radiation (CMBR) and applied as cosmological standard rod (e.g. Blake et al 2012; Padmanabhan et al. 2012; Anderson et al. 2012).

iii. Other test include the redshift-angular size test, galaxy cluster gas mass fraction, strong gravitational lensing test and structure formation test, these test generally constrain cosmological parameters as a function of redshift (see review by Samushia & Rastra 2006).

Several decades have passed since Hubble published the correlation between distances to galaxies and their expansion velocities (Hubble 1929), but establishing an accurate cosmological distance scale and value for the Hubble constant ($H_0$) have proved challenging. The value of $H_0$ has evolved from $H_0 = 500 \, kms^{-1} \, Mpc^{-1}$ recorded by Hubble (1929) to a well-known range of range of $H_0 = 50 - 100 \, kms^{-1} \, Mpc^{-1}$ (Assis et al. 2009). The Hubble Space Telescope (HST) gave a more specific value of $H_0 = 70 - 80 \pm 10\%$ (Freedman et al. 2001), while the WAMP data give $H_0 = 72 \pm 5 \, kms^{-1} \, Mpc^{-1}, \Omega_m = 0.3$, and $\Omega_\Lambda = 0.7$ (Spergel et al. 2003), where $\Omega_m$ is relative matter density (both luminous and dark matter) and $\Omega_\Lambda$ is the relative dark energy density.

In this article, we use the dependence of the observed distance modulus $(m - M)$ on redshift $(z)$ and assumed values of the deceleration parameter $(q_0)$ to constrain the Hubble constant using data obtained from NED and possibly trace the evolution of the Hubble parameter as a function of redshift $(H(z))$. In Section 2, we establish the relation between distance modulus, redshift and deceleration parameter. The data is analyzed in Section 3 and we conclude in Section 4.

## 2  THEORETICAL RELATIONSHIPS

The evolutionary trend of the universe has been modeled by using the density parameter $(\Omega)$, determine by the density of the universe, the deceleration parameter $(q_0)$ and the Hubble constant $(H_0)$. The Hubble constant is an important parameter in cosmology as it not only determines its expansion rate, it also set limit to the possible age, critical density and size of the observable universe. The velocity $(v)$ of the expansion of the universe has been defined by Hubble (1925) as

$$v(t) = H_0(t)d, \qquad (1)$$

where $d$ is the radius of the expanding universe. The size of the universe is unknown, yet it undergoes expansion or contraction, thus, the evolution of the universe can be express in terms of cosmic scale factor $(a(t))$ as

$$a(t) = \frac{R(t_e)}{R_0(t_0)}. \qquad (2)$$

where $R(t_e)$ is the expansion factor at an earlier time $t_e$ and $R_0(t_0)$ is the expansion factor at a later time $t_0$. In terms of the cosmic expansion factor $(R_0)$, at the time $(t_0)$, the Hubble constant is given by (e.g. Roos 2003)

$$H_0(t) = \frac{\dot{R}_0(t)}{R_0(t_0)} \rightarrow H(t) = \frac{\dot{R}(t)}{R(t)} = \frac{\dot{a}(t)}{a(t)}, \qquad (3)$$

where $\dot{R}_0(t)$ is the first time derivative of $R_0$. The cosmic scale factor affects all distances, thus, in terms of the wavelength $(\lambda_e)$ of a photon emitted at time of emission $(t_e)$ and observed at another time $(t_0)$, will be the ratio of the rates of expansion can be written as (e.g. Roos 2003)

$$\frac{\lambda_0}{\lambda_e} = \frac{R_0(t_0)}{R(t_e)} = [a(t)]^{-1}. \qquad (4)$$

where $\lambda_0$ is the wavelength of the observed photon at time $(t_0)$. The cosmological redshift $(z)$ is usually given by (e.g. Roos 2003)

$$z = \frac{\lambda_0 - \lambda_e}{\lambda_e} = (t_0 - t_e)H_0. \qquad (5)$$

Allowing for time evolution of the expansion factor $(R(t))$, we expand it using Taylor series (Roos 2003), for $t$ in general we can write

$$R(t) \approx R_0 + \dot{R}_0(t - t_0) + \frac{1}{2}\ddot{R}_0(t - t_0)^2. \qquad (6)$$

Making use of the definition of equation (3), equation (6) implies that to a second order expansion, the cosmic scale factor can be written as

$$a(t) = 1 - H_0(t_0 - t) + \frac{1}{2}\dot{H}_0(t_0 - t)^2 \qquad (7)$$

From equation (3), we can write

$$\dot{R}_0(t_0 - t_e) = H_0 R_0(t_0 - t_e). \qquad (8)$$

The deceleration parameter $(q)$ which measure the rate of slowing down of the expansion factor is defined (e.g. Roos 2003) by

$$q = -\frac{a\ddot{a}}{\dot{a}^2} = -\frac{\ddot{a}}{aH^2}. \qquad (9)$$

Makin use of equations (3), (4), (7) and (9) in equation (5), the cosmological redshift can be expressed as

$$z(t) + 1 = \left[1 - \left\{H_0(t_0 - t_e) + \frac{1}{2}q_0 H_0^2 (t_0 - t_e)^2\right\}\right]^{-1}. \tag{10}$$

where $\dot{H}_0 = \frac{\ddot{a}_0}{a_0} = -q_0 H_0^2$. Letting $x = \left[H_0(t_0 - t_e) + \frac{1}{2}q_0 H_0^2 (t_0 - t_e)^2\right]$ and making use of the series expansion $(1-x)^{-1} \approx 1 + x + x^2$, equation (10) can be approximated to

$$z(t) = H_0(t_0 - t_e) + (1 + \frac{1}{2}q_0)H_0^2(t_0 - t_e)^2. \tag{11}$$

In obtaining equation (11), we made use of terms only to the second order in $t$. Inverting equation (11) and making use of equation (3), we have $H_0$ in terms of redshift ($z$) as

$$H_0(t_0 - t_e) = z - \left(1 + \frac{1}{2}q_0\right)z^2. \tag{12}$$

For an isotropic and homogeneous universe, the Robertson–Walker metric in Minkowiski spacetime best describe the geometry of space and it is given as (Camenzind 2009)

$$ds^2 = c^2 dt^2 - R^2(t)\left(\frac{dr^2}{1-kr^2} + r^2 d\theta^2 + r^2 \sin^2 d\varphi^2\right), \tag{13}$$

where $ds$ is the invariant line element, $r$ is the radial distance, $k$ is the curvature parameter, $\theta$ and $\varphi$ are spherical coordinate points (polar and azimuthal angle respectively). In such homogenous isotropic universe, photons propagates along null geodesic given by $ds^2 = 0$, and along the line of sight of an observer, where $\theta$ and $\varphi$ are assumed constant, $d\theta = d\varphi = 0$. Thus, equation (13) becomes

$$\frac{cdt}{R(t)} = \frac{dr}{\sqrt{1-kr^2}} \tag{14}$$

Integrating equation (14) from the source to the observer, we let the time at source be $t_e$, (emission time) and that at the observer be $t_0$ (present time), for a flat universe ($k = 0$ Roos (2003)), we have

$$\int_{t_e}^{t_0} \frac{cdt}{R(t)} = \int_0^r dr = r \tag{15}$$

But for photons that were emitted in the past $R(t) = R(t_e)$ (Ryden 2006) and multiplying through with $R(t_0)$, equation (15) becomes

$$R(t_0)c \int_{t_e}^{t_0} \frac{cdt}{R(t)} = \int_{t_e}^{t_0} \frac{dt}{a(t)} = R(t_0)r \tag{16}$$

Substituting $a(t) \approx 1 - H_0(t_0 - t_e)$ (see equation 7) in equation (16) and defining $R(t_0)r$ as the proper distance to any given galaxy centered on the observer. Evaluating equation (16) using equation (12) gives (see Roos 2003)

$$d_p = c \int_{t_e}^{t_0}[1 + H_0(t_0 - t_e)]dt = \frac{cz}{H_0}\left[1 - \frac{1}{2}(1 + q_0)z\right] \qquad (17)$$

The first term on the right hand side of equation (17) gives the Hubble linear law, while the second term measures the deviation from linearity to lowest order depending on the value of $q_0$.

The luminosity distance $(d_L)$ to a galaxy of absolute luminosity $(L)$ with observed brightness $(B_o)$ is given by (e.g. Roos 2003)

$$d_L = \sqrt{\frac{L}{4\pi B_o}} \qquad (18)$$

For an expanding universe parameterized by the cosmic scale factor $(a(t))$, in which photons are redshifted and suffer from energy effect, if the apparent brightness of a galaxy is $B_a$, then its proper distance is given by (e.g. Roos 2003)

$$d_p = \frac{1}{1+z}\sqrt{\frac{L}{4\pi B_a}} \qquad (19)$$

Equating $B_a = B_o$, implies that $d_L = (1 + z)d_P$. Thus, the luminosity distance in terms of redshift to a first order approximation (small values of z) is given by

$$d_L = \frac{cz}{H_0}\left[1 + \frac{z}{2}(1 - q_0)\right] \qquad (20)$$

In terms of distance modulus $(m - M)$, the luminosity distance to a source is given by (e.g. Hogg 1998)

$$m - M = 5\log\left(\frac{d_L}{10(pc)}\right) \qquad (21)$$

Substituting equation (20) into equation (21), we have

$$m - M = 42.3856 - 5\log h + f(z) \qquad (22)$$

where $f(z) = 5\log z + 5\log\left[1 + \frac{z}{2}(1 - q_0)\right]$ and incorporates some form of non-linear evolution of luminosity distance with redshift with $H_0 = 100h\ kms^{-1}\ Mpc^{-1}$ (e.g. Roos 2003) and in S.I is given by $H_0 = 3.241h\ x10^{-18}\ s^{-1}$. For different energy densities of the universe characterized by deceleration parameter as $q_0 = 0$, $q_0 = 1.0$ and $q_0 = -1.0$ (e.g. Lima 2008) the redshift function in equation (22) reduces, respectively, to

$$f(z) = 5\log z + 5\log\left(1 + \frac{z}{2}\right) \qquad (23a)$$

$$f(z) = 5\log z \qquad (23b)$$

$$f(z) = 5\log z + 5\log(1 + z) \qquad (23c)$$

Equation (20) provides a way to constrain the Hubble parameter ($h$) for a given sample of sources with observed redshift and distance modulus.

## 3. ANALYSIS AND RESULTS

The data used in this analysis were sourced from the NASA Extragalactic Database (NED). We selected sources with observed distance modulus ($m - M$) and redshift ($z$). The sample consists of 11 585 sources which is spread over $0.00016 \leq z \leq 8.26$. Due to large volume of data, we binned the sources into various redshift bin widths to enable fair representation of sources in each bin. The choice of the binning range is arbitrary.

Using the whole sources in our sample, the median value of the Hubble parameter we estimated is $h = 0.7 \pm 0.3$ for $q_0 = 0$, $h = 0.6 \pm 0.3$, for $q_0 = 1$ and $h = 0.8 \pm 0.3$, for $q_0 = -1$. Using these values of Hubble parameter, the corresponding age ($\tau$) and size ($\Re$) of the universe we estimated are: for $q_0 = 0, \tau = 15 \pm 1\ Gyrs$, $\Re = (5 \pm 2) \times 10^3\ Mpc$, for $q_0 = 1, \tau = 18 \pm 1\ Gyrs$, $\Re = (6 \pm 2) \times 10^3\ Mpc$ and for $q_0 = -1, \tau = 13 \pm 1\ Gyrs$, $\Re = (4 \pm 2) \times 10^3\ Mpc$. The quoted errors are standard deviation from the average values of the whole sources in our sample. The binning ranges, the mean, median values of each bin size and the estimated Hubble parameter and estimated age and size of the universe from the estimated Hubble parameter based on equation (20) are shown in Table 1. Using the median value of each bin, we plot $m - M$ against $f(z)$ for different values of the deceleration parameter. The plots are shown in figures 1-3. The error bars indicate the errors in the distance modulus ($m - M$).

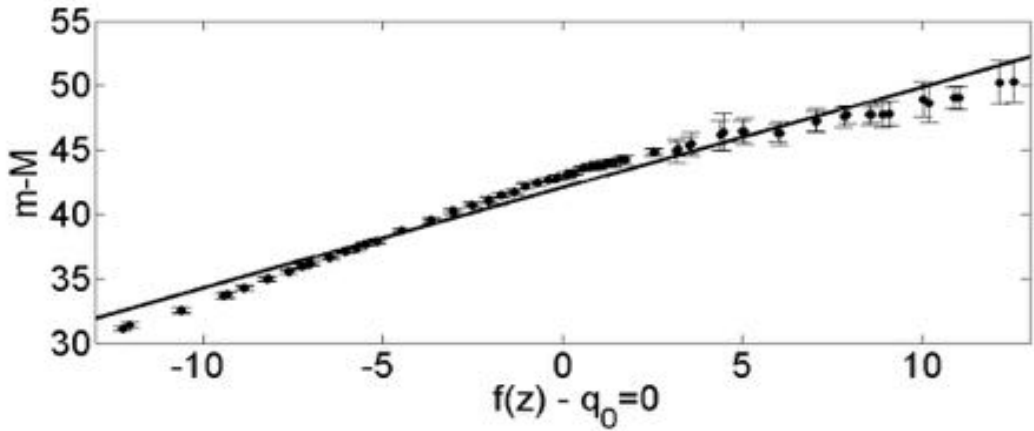

Fig 1. The plot of $m - M$ against $f(z)$ for $q_0 = 0$

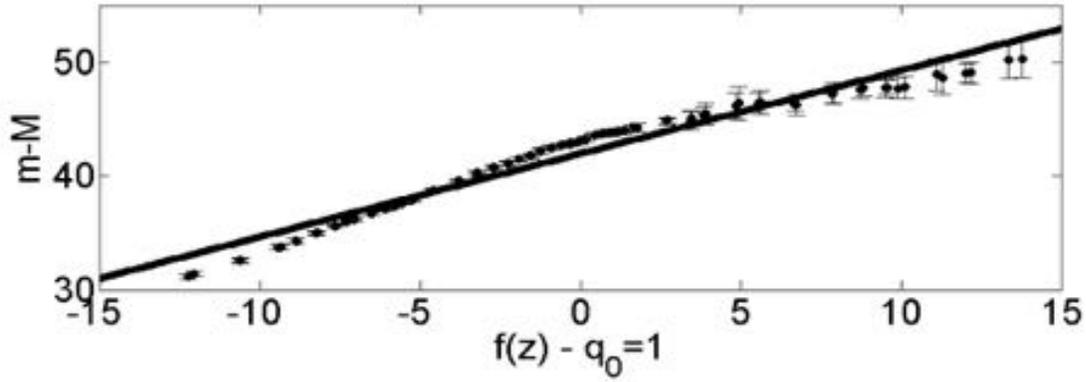

Fig 2. The plot of $m - M$ against $f(z)$ for $q_0 = 1$

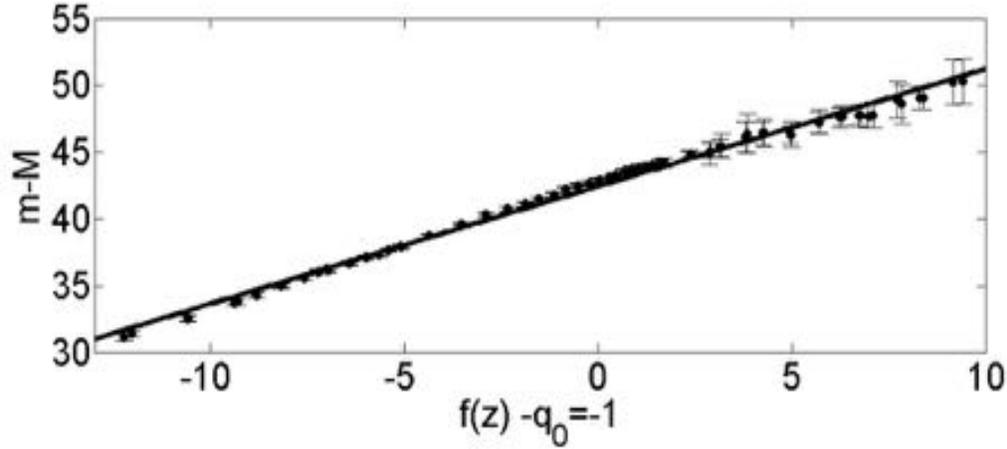

Fig 3. The plot of $m - M$ against $f(z)$ for $q_0 = -1$

Linear regression fit to the plots give: (a) for $q_0 = 0$ we have $m - M = (0.95 \pm 0.33)f(z) + 42.99 \pm 1.87 \rightarrow 0.63 \leq h \leq 0.87, r = 0.95$ while the age and size limits are $11.3 \leq \tau(Gyr) \leq 15.5$ and $3.45 \times 10^3 \leq \Re(Mpc) \leq 4.77 \times 10^3$ respectively; (b) for $q_0 = 1$ we have $m - M = (1.13 \pm 0.28)f(z) + 44.20 \pm 1.73 \rightarrow 0.51 \leq h \leq 0.73, r = 0.95$ with age and size limits as $13.4 \leq \tau(Gyr) \leq 19.2$ and $4.11 \times 10^3 \leq \Re(Mpc) \leq 5.88 \times 10^3$ respectively; (c) for $q_0 = -1$ we have $m - M = (0.88 \pm 0.39)f(z) + 42.47 \pm 2.04 \rightarrow 0.66 \leq h \leq 1.30, r = 0.98$, the age and size limit are estimated as $7.53 \leq \tau(Gyr) \leq 14.7$ and $2.31 \times 10^3 \leq \Re(Mpc) \leq 4.52 \times 10^3$. The slopes of the plots approximate the theoretical value with strong correlation ($r$ is the correlation coefficient)

## 4. DISCUSSION AND CONCLUSION

The value of Hubble constant ($H_0$) we estimated lies in the range of $60 \pm 25\ kms^{-1} Mpc^{-1} \leq H_0 \leq 98 \pm 43\ kms^{-1} Mpc^{-1}$ with an average value of $H_0 = 67 \pm 22 - 96 \pm 29\ kms^{-1} Mpc^{-1}$. The mean value of $H_0$ we estimated is in reasonable agreement with the HST value of $H_0 = 70 - 82 \pm 10\% \; kms^{-1}Mpc^{-1}$ recorded by Freedman (2001). Our result in general is also in agreement with results obtained from other works in literature e.g. $H_0 = 50 - 100 \; kms^{-1}Mpc^{-1}$ by Assis et al. (2009), $H_0 = 72 \pm 5 \; kms^{-1}Mpc^{-1}$ from WMAP data by Spergel et al. (2003), $H_0 = 74.1^{+7.9}_{-7.1} \; kms^{-1}Mpc^{-1}$ by Aviles et al. (2012), $H_0 = 79 \pm 30\% \; kms^{-1}Mpc^{-1}$ by Blinnikov et al. (2012), $H_0 = 74 \pm 2.5 \; kms^{-1}Mpc^{-1}$ by Lima et al. (2012) similar to $H_0 = 74 \pm 2.5 \; kms^{-1}Mpc^{-1}$ by Freedman et al. (2012) and $H_0 = 68.9 \pm 7.1 \; kms^{-1}Mpc^{-1}$ by Reid et al. (2013). The limit of the range $60 kms^{-1}Mpc^{-1} \leq H_0 \leq 220 \; kms^{-1}Mpc^{-1}$ is also in agreement with $H_0 = 60 - 220 \; kms^{-1}Mpc^{-1}$ obtained by Farooq & Rastra (2012).

In conclusion, using the observed distance modulus and redshift, we estimated Hubble parameter, for different assumed energy densities of the universe represented by the deceleration parameter ($q_0$). The age of the universe estimated from the Hubble parameter suggests that the universe is $\sim 11.4 - 19.3 \; Gyr$ old with the possible limit to the edge of the observable universe of $d \sim 3450 - 5880 \; Mpc$.

Table 1: Estimated Values of Hubble Parameter, Age and Size of Observable Universe for Different bin Values of Redshift

| $z$ | $h$ ($q_0 = 0$) | $h$ ($q_0 = 1$) | $h$ ($q_0 = -1$) | $\tau$ ($q_0 = 0$) $\times 10^9$ yrs | $\tau$ ($q_0 = 1$) $\times 10^9$ yrs | $\tau$ ($q_0 = -1$) $\times 10^9$ yrs | $\Re$ ($q_0 = 0$) $\times 10^3$ Mpc | $\Re$ ($q_0 = 1$) $\times 10^3$ Mpc | $\Re$ ($q_0 = 0$) $\times 10^3$ Mpc |
|---|---|---|---|---|---|---|---|---|---|
| 0.004 | 0.64 | 0.64 | 0.64 | 15.34 | 15.36 | 15.31 | 4.70 | 4.71 | 4.70 |
| 0.008 | 0.71 | 0.70 | 0.71 | 13.85 | 13.90 | 13.80 | 4.25 | 4.27 | 4.23 |
| 0.013 | 0.72 | 0.71 | 0.72 | 13.64 | 13.73 | 13.55 | 4.18 | 4.21 | 4.16 |
| 0.017 | 0.72 | 0.71 | 0.72 | 13.62 | 13.73 | 13.50 | 4.18 | 4.21 | 4.14 |
| 0.022 | 0.70 | 0.69 | 0.71 | 14.02 | 14.18 | 13.87 | 4.30 | 4.35 | 4.25 |
| 0.029 | 0.70 | 0.69 | 0.71 | 14.04 | 14.25 | 13.84 | 4.31 | 4.37 | 4.25 |
| 0.035 | 0.68 | 0.67 | 0.69 | 14.44 | 14.69 | 14.20 | 4.43 | 4.51 | 4.36 |
| 0.039 | 0.70 | 0.69 | 0.72 | 13.92 | 14.19 | 13.66 | 4.27 | 4.35 | 4.19 |
| 0.049 | 0.71 | 0.69 | 0.72 | 13.83 | 14.17 | 13.51 | 4.24 | 4.35 | 4.14 |
| 0.060 | 0.71 | 0.69 | 0.73 | 13.76 | 14.17 | 13.37 | 4.22 | 4.35 | 4.10 |
| 0.069 | 0.75 | 0.72 | 0.77 | 13.05 | 13.51 | 12.63 | 4.00 | 4.14 | 3.88 |
| 0.077 | 0.71 | 0.68 | 0.73 | 13.80 | 14.34 | 13.31 | 4.23 | 4.40 | 4.08 |
| 0.089 | 0.73 | 0.70 | 0.76 | 13.36 | 13.95 | 12.82 | 4.10 | 4.28 | 3.93 |
| 0.120 | 0.68 | 0.64 | 0.72 | 14.42 | 15.28 | 13.65 | 4.42 | 4.69 | 4.19 |
| 0.169 | 0.70 | 0.65 | 0.76 | 13.96 | 15.14 | 12.95 | 4.28 | 4.65 | 3.97 |
| 0.218 | 0.68 | 0.62 | 0.75 | 14.29 | 15.85 | 13.01 | 4.38 | 4.86 | 3.99 |
| 0.270 | 0.66 | 0.58 | 0.74 | 14.76 | 16.75 | 13.20 | 4.53 | 5.14 | 4.05 |
| 0.320 | 0.69 | 0.59 | 0.78 | 14.25 | 16.53 | 12.52 | 4.37 | 5.07 | 3.84 |
| 0.366 | 0.66 | 0.56 | 0.76 | 14.89 | 17.61 | 12.90 | 4.57 | 5.40 | 3.96 |
| 0.417 | 0.68 | 0.57 | 0.80 | 14.28 | 17.26 | 12.18 | 4.38 | 5.29 | 3.74 |
| 0.464 | 0.62 | 0.50 | 0.74 | 15.79 | 19.46 | 13.29 | 4.85 | 5.97 | 4.08 |
| 0.516 | 0.63 | 0.50 | 0.76 | 15.50 | 19.49 | 12.86 | 4.75 | 5.98 | 3.95 |
| 0.571 | 0.64 | 0.50 | 0.79 | 15.17 | 19.49 | 12.41 | 4.65 | 5.98 | 3.81 |
| 0.620 | 0.67 | 0.51 | 0.82 | 14.68 | 19.22 | 11.87 | 4.50 | 5.90 | 3.64 |
| 0.674 | 0.67 | 0.50 | 0.84 | 14.52 | 19.42 | 11.60 | 4.45 | 5.96 | 3.56 |
| 0.716 | 0.66 | 0.48 | 0.83 | 14.86 | 20.18 | 11.76 | 4.56 | 6.19 | 3.61 |
| 0.769 | 0.62 | 0.45 | 0.79 | 15.78 | 21.84 | 12.35 | 4.84 | 6.70 | 3.79 |
| 0.821 | 0.63 | 0.45 | 0.82 | 15.47 | 21.82 | 11.98 | 4.75 | 6.69 | 3.68 |

| 0.868 | 0.64 | 0.45 | 0.84 | 15.18 | 21.76 | 11.65 | 4.66 | 6.68 | 3.57 |
| 0.921 | 0.66 | 0.45 | 0.87 | 14.85 | 21.68 | 11.29 | 4.55 | 6.65 | 3.46 |
| 0.967 | 0.70 | 0.47 | 0.92 | 14.03 | 20.81 | 10.58 | 4.30 | 6.38 | 3.25 |
| 1.063 | 0.70 | 0.46 | 0.94 | 14.00 | 21.45 | 10.40 | 4.30 | 6.58 | 3.19 |
| 1.290 | 0.70 | 0.42 | 0.97 | 14.02 | 23.07 | 10.07 | 4.30 | 7.08 | 3.09 |
| 1.499 | 0.81 | 0.46 | 1.16 | 12.09 | 21.14 | 8.46 | 3.71 | 6.49 | 2.60 |
| 1.638 | 0.72 | 0.40 | 1.04 | 13.60 | 24.73 | 9.37 | 4.17 | 7.59 | 2.88 |
| 1.946 | 0.68 | 0.34 | 1.01 | 14.42 | 28.45 | 9.66 | 4.42 | 8.73 | 2.96 |
| 2.230 | 0.72 | 0.34 | 1.10 | 13.52 | 28.60 | 8.86 | 4.15 | 8.77 | 2.72 |
| 2.694 | 1.05 | 0.45 | 1.65 | 9.33 | 21.90 | 5.93 | 2.86 | 6.72 | 1.82 |
| 3.257 | 0.89 | 0.34 | 1.44 | 10.98 | 28.85 | 6.78 | 3.37 | 8.85 | 2.08 |
| 3.730 | 0.97 | 0.34 | 1.60 | 10.08 | 28.89 | 6.11 | 3.09 | 8.86 | 1.87 |
| 4.230 | 1.11 | 0.36 | 1.87 | 8.79 | 27.38 | 5.23 | 2.70 | 8.40 | 1.61 |
| 4.654 | 1.28 | 0.38 | 2.17 | 7.65 | 25.45 | 4.50 | 2.35 | 7.81 | 1.38 |
| 5.437 | 1.00 | 0.27 | 1.73 | 9.78 | 36.36 | 5.65 | 3.00 | 11.15 | 1.73 |
| 6.419 | 1.24 | 0.29 | 2.19 | 7.89 | 33.20 | 4.47 | 2.42 | 10.18 | 1.37 |
| 7.738 | 1.02 | 0.21 | 1.82 | 9.63 | 46.91 | 5.37 | 2.96 | 14.39 | 1.65 |